\documentclass[namedreferences]{solarphysics}
\usepackage[hyperref,optionalrh,solaromanenum]{spr-sola-addons} 
\usepackage{graphicx}                    
\usepackage{amssymb}                     
\usepackage{color}                       
\usepackage{breakurl}                    

\newcommand{\revone}{\bf \color{blue}}
\newcommand{\degree}{\ensuremath{^\circ}}

\newcommand{\soho}{{\em SOHO{}}}

\newcommand{\stereo}{{\em STEREO{}}}
\newcommand{\sdo}{{\em SDO{}}}

\runningauthor{S. W. McIntosh {\em et~al.}}

\begin{document}

\begin{opening}
  \title{What the Sudden Death of Solar Cycles Can Tell us About the Nature of the Solar Interior}

  \author[addressref={1},corref,email={mscott@ucar.edu}]{\inits{S.W.}\fnm{Scott W.}~\lnm{McIntosh}~\orcid{0000-0002-7369-1776}}
  \author[addressref={2,3}]{\inits{R.J.}\fnm{Robert J.}~\lnm{Leamon}~\orcid{0000-0002-6811-5862}}
  \author[addressref={1}]{\inits{R.E.}\fnm{Ricky}~\lnm{Egeland}~\orcid{0000-0002-4996-0753}}
  \author[addressref={1}]{\inits{M.}\fnm{Mausumi}~\lnm{Dikpati}~\orcid{0000-0002-2227-0488}}
  \author[addressref={1}]{\inits{Y.}\fnm{Yuhong}~\lnm{Fan}~\orcid{0000-0003-1027-0795}}
  \author[addressref={1}]{\inits{M.}\fnm{Matthias}~\lnm{Rempel}~\orcid{0000-0001-5850-3119}}
  \address[id={1}]{National Center for Atmospheric Research, High Altitude Observatory, P.O. Box 3000, Boulder, CO~80307, USA}
  \address[id={2}]{University of Maryland, Department of Astronomy, College Park, MD 20742, USA}
  \address[id={3}]{NASA Goddard Space Flight Center, Code 672, Greenbelt, MD 20771, USA.}
  
  \begin{abstract}
    We observe the abrupt end of solar activity cycles at the Sun's Equator by combining almost 140 years of observations from ground and space. These ``terminator'' events appear to be very closely related to the onset of magnetic activity belonging to the next Solar Cycle at mid-latitudes and the polar-reversal process at high latitudes. Using multi-scale tracers of solar activity we examine the timing of these events in relation to the excitation of new activity and find that the time taken for the solar plasma to communicate this transition is of the order of one solar rotation -- but could be shorter. Utilizing uniquely comprehensive solar observations from the {\it Solar Terrestrial Relations Observatory} ({\it STEREO}) and {\it Solar Dynamics Observatory} ({\it SDO}) we see that this transitional event is strongly longitudinal in nature. Combined, these characteristics suggest that information is communicated through the solar interior rapidly. A range of possibilities exist to explain such behavior: for example gravity waves on the solar tachocline, or that the magnetic fields present in the Sun's convection zone could be very large, with a poloidal field strengths reaching 50~kG -- considerably larger than conventional explorations of solar and stellar dynamos estimate. Regardless of the mechanism responsible, the rapid timescales demonstrated by the Sun's global magnetic field reconfiguration present strong constraints on first-principles numerical simulations of the solar interior and, by extension, other stars.
  \end{abstract}

\end{opening}

\section{Introduction}
Understanding the nature of the solar dynamo, the mechanism by which the Sun's magnetic field is generated, sustained, and repeatedly cycles, remains one of the most challenging matters in astrophysics \citep{2010LRSP....7....1H, 2010LRSP....7....3C}. Over the past fifty years, models attempting to reproduce the key ingredients of solar variability have grown in sophistication from observationally motivated empirical models \citep{1961ApJ...133..572B,1969ApJ...156....1L} to fully three-dimensional magnetohydrodynamic numerical simulations \citep{2011ApJ...735...46R,2014ApJ...789...35F,2016Sci...351.1427H} with the common top-level goal of reproducing the spatio-temporal variation of sunspots over decades. Levels of success in such modeling efforts have been mixed \citep{2010LRSP....7....3C}, as many of the key physical properties of the deep solar interior, where the large-scale dynamo action is believed to take place, remain largely unknown and are extremely difficult to observe.

Extreme Ultraviolet (EUV) Bright Points (BPs) are an ubiquitous feature of the solar corona and, since the Skylab era, research has focused on their origin and diagnostic potential, hinting at their ties to the magnetism of the deep solar convection zone \citep{1974ApJ...189L..93G, 1978ApJ...219L..55G, 1980RSPTA.297..595G, 1988Natur.333..748W}. Recent research has solidified their connection to the scales of rotationally driven ``giant-cell" convection \citep{2013Sci...342.1217H, 2014ApJ...784L..32M}. \cite{2014ApJ...792...12M} (hereafter M2014) demonstrated that monitoring the latitudinal progression of the number density of BPs with time \citep{2005SoPh..228..285M} permitted the identification and tracking of magnetic activity bands that migrate from latitudes around 55\degree{} to the Equator over a period of around 20 years. These bands of enhanced BP activity presented an extension in space and time of the sunspot activity and migration pattern, known as the ``butterfly pattern'', which has been observed for over a century \citep{Maun04} and has been considered as the pattern central to the Sun's dynamo mystery \citep{2010LRSP....7....3C}. As part of their investigation, M2014 noticed an abrupt reduction in the equatorial BP density that occurred in conjunction with the very rapid growth of BP density at mid\--latitudes ($\approx$35\degree{} latitude). The rapid growth of BP density at mid-latitudes appeared to be occurring at same time as the rapid increase in the number of sunspots being produced at the same latitudes -- those spots belonged to the new Solar Cycle. M2014 deduced that the switch-over, transition, or annihilation of the oppositely polarized magnetic bands at the Equator was a required event to trigger the emergence of new cycle spots at mid\--latitudes. These events were deemed to mark the termination of the old magnetic activity bands at the solar Equator, and hence were the transition points from Solar Cycle 22 to 23 (in 1997) and from Solar Cycle 23 to 24 (in 2011). M2014 inferred that the enhanced BP- density bands were intrinsically tied to the Sun's 22\--year magnetic-polarity cycle \citep{1925ApJ....62..270H} and that the interaction of those activity bands was responsible for the modulation of the sunspot pattern (see below). The ``termination'' events, or ``terminators,'' that are described in M2014 took place over a time span of around one solar rotation -- a remarkably short timescale for a process inside a star. Exploring their duration in more detail motivates this article; what might we learn about the state of the solar interior by considering these termination events?

In the following we will introduce and illustrate certain concepts from M2014 and demonstrate the appearance of terminators in a host of solar features, measures, and standard indices going back nearly 140 years. Culminating in the contemporary era, we revisit M2014's terminators in the 22-year long record of coronal images taken by the Extreme Ultraviolet Imaging Telescope \citep[EIT:][]{1995SoPh..162..291D} onboard the Solar and Heliospheric Observatory (\soho), the Extreme Ultraviolet Imagers \citep[EUVI:][]{2008SSRv..136...67H} onboard the twin \stereo{} spacecraft and the Atmospheric Imaging Assembly \citep[AIA:][]{2012SoPh..275...17L} onboard the \sdo{} spacecraft in the 195 and 193\AA{} broadband channels. While the contemporary dataset permits a more precise timing and longitudinal exploration of the most recent terminator, examination of the entire record offers possible insight into the physics of the solar interior, how it constructs the patterns of solar activity, and the possible importance of terminators in the latter.

\section{Observations \& Analysis}

\subsection{Terminators in the Sunspot Record}
\begin{figure}[!ht]
\centering
\includegraphics[width=1.00\linewidth]{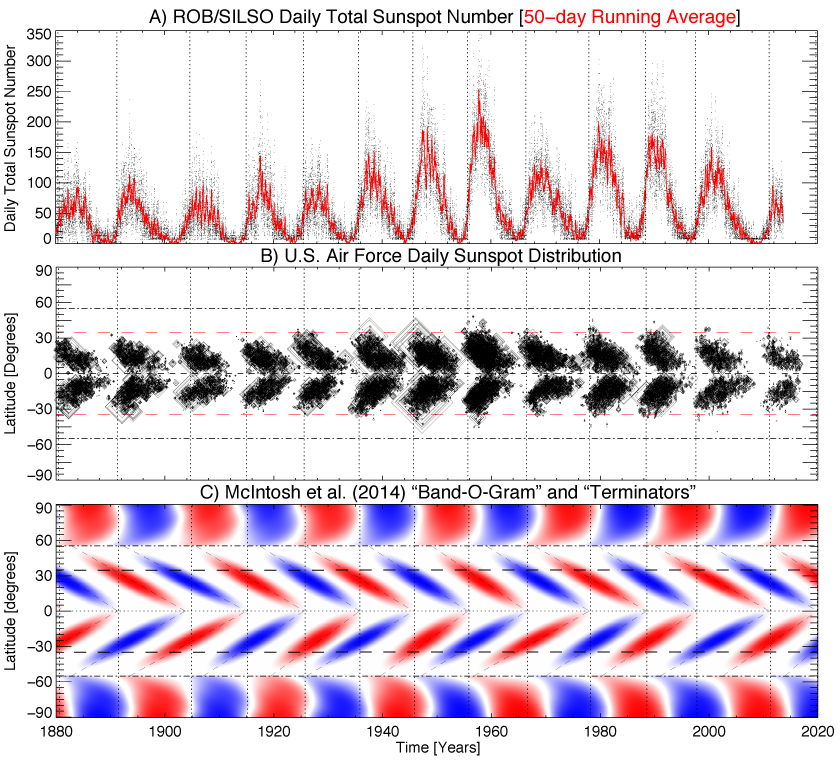}
\caption{Solar activity over the past 140 years (1860\--2020) through the observed variation in sunspot number, their distribution with latitude, and the M2014 ``band-o-gram''. In {\bf Panel A}, we show the daily total sunspot number with a 50-day running average {\it overplotted in red}. The 11-year quasi-cyclic variability is clear. In {\bf Panel B}, the sunspot locations are shown, and their relative extent is indicated by the size of the symbol. Note that the sunspot butterfly wings grow abruptly at latitudes largely less than 35\degree{}. {\bf Panel C} shows the ``band-o-gram,'' a schematic depiction of solar magnetic activity, that is constructed from the information in Panel B and Table~\ref{T1}. In each panel the dotted vertical lines are the termination events, or terminators, measured by M2014. In the two lower panels, for reference, we show {\it dashed} and {\it dot\--dashed} lines at 35\degree{} and 55\degree{} latitude, respectively.}
\label{fX}
\end{figure}

Figure~\ref{fX}A shows the familiar, quasi-cyclic number of sunspots present on the Sun since 1860. Despite decades of research, this coarsest measure of solar variability remains a quintessential enigma of astrophysics  \citep{2010LRSP....7....1H}. Figure~\ref{fX}B adds to the complexity of sunspot evolution by demonstrating the evolution of spots in latitude and time. This plotting method results in the ``butterfly diagram'' of the Maunders \citep{Maun04}, because the cross-equatorial group of spots presented in each solar cycle resembles the outstretched wings of a butterfly. On each of these panels we have drawn dashed-vertical lines that represent the terminator events. These dashed lines were presented by M2014 as the times when the area of the sunspots on the disk became greater than 100 millionths of the Sun's disk area following sunspot minimum. This designation in M2014 was motivated by their noticing the aforementioned terminator events in EUV BP density of 1997 and 2011 that were coincident with a very rapid onset of sunspot appearance and area in both solar hemispheres, on both occasions. In an effort to extend their analysis to times before EUV BPs were available, and also to compare with other recent observables, their analysis used a threshold, 100 millionths of the solar disk area covered by sunspots, to identify possible termination events and placed them in Table~1 of that article. The dashed lines present in Figure~\ref{fX} are provided again in this work's Table~\ref{T1} along with other climatalogical measures. Notice that the terminator events fall at neither solar maximum (the peak of sunspot activity), nor at solar minimum (the bottom of sunspot activity), instead appearing, apparently, just after the start of the Solar Cycles: the ``ascending phase''.  

\setlength{\tabcolsep}{3pt}
\begin{table}
 \tiny
 \caption{Features required to build a band-o-gram, converted from Table~1 of M2014. From left to right we provide the Solar Cycle number; the start times of the bands in each hemispheric (the time of the previous hemispheric maxima); the difference between hemispheric maxima [$\delta = T_{N} - T_{S}$] where red indicates north leading and blue for the south; the time between hemispheric maxima of the same magnetic polarity [$\delta_{N}$ and $\delta_{S}$]; the terminator; the time elapsed since last terminator [$\Delta$]; and the hemispheric high\--low latitude transit time [$\tau$] for each hemisphere. As a reminder, $T_{N,S}$ is hemispheric sunspot maximum and, as per M2014, is also the first incursion of the band that becomes the upcoming Solar Cycle.}\label{T1}
  \begin{tabular}{cccccc|cc|ccc}
    \hline
	Cycle $\#$ & T$_N$ [yr] & T$_S$ [yr] & $\delta$ [yr] & 
    $\delta_{N}$ [yr] & $\delta_{S}$ [yr] & Terminator [yr] & 
    $\Delta$ [yr] & $\tau_{N}$ [yr] & $\tau_{S}$ [yr] \\
	\hline
    12 & $\ldots$ & $\ldots$ & $\ldots$ & $\ldots$ & $\ldots$ & 1891.30 & $\ldots$ & $\ldots$ & $\ldots$  \\    
    13 & 1884.00 & 1883.83 & \textcolor{blue}{0.17} & $\ldots$ & $\ldots$ & 1904.75 & 13.45 & 20.75 & 20.92 \\
    14 & 1892.50 & 1893.58 & \textcolor{red}{-1.08} & $\ldots$ & $\ldots$ & 1915.05 & 10.30 & 22.55 & 21.47 \\
    15 & 1905.75 & 1907.08 & \textcolor{red}{-1.33} & 21.75 & 23.25 & 1925.67 & 10.62 & 19.92 & 18.59 \\
    16 & 1917.58 & 1919.50 & \textcolor{red}{-1.92} & 25.08 & 25.92 & 1935.75 & 10.08 & 18.17 & 16.25 \\
	17 & 1925.92 & 1926.08 & \textcolor{red}{-0.16} & 20.17 & 19.00 & 1945.75 & 10.00 & 19.83 & 19.67 \\
    18 & 1937.50 & 1939.67 & \textcolor{red}{-2.17} & 19.92 & 20.17 & 1955.75 & 10.00 & 18.25 & 16.08 \\
    19 & 1949.17 & 1947.17 & \textcolor{blue}{2.00} & 23.25 & 21.08 & 1966.50 & 10.75 & 17.33 & 19.33 \\
	20 & 1959.08 & 1956.83 & \textcolor{blue}{2.25} & 21.58 & 17.17 & 1978.00 & 11.50 & 18.92 & 21.17 \\
    21 & 1968.00 & 1970.08 & \textcolor{red}{-2.08} & 18.83 & 22.92 & 1988.50 & 10.00 & 20.50 & 18.42 \\
    22 & 1979.67 & 1980.33 & \textcolor{red}{-0.66} & 20.58 & 23.50 & 1997.75 &  9.25 & 18.08 & 17.42 \\
	23 & 1989.08 & 1991.08 & \textcolor{red}{-2.00} & 21.08 & 21.00 & 2011.20 & 13.45 & 22.12 & 20.12 \\
    24 & 2000.50 & 2002.58 & \textcolor{red}{-2.08} & 20.83 & 22.25 & $\ldots$ & $\ldots$ & $\ldots$ & $\ldots$ \\
    25 & 2011.80 & 2013.95 & \textcolor{red}{-2.15} & 22.72 & 22.87 & $\ldots$ & $\ldots$ & $\ldots$ & $\ldots$ \\
	\hline
 Means &  &  &  & 21.44 & 21.74 & & 10.85 & 19.67 & 19.04 \\
 Std. Dev. &  &  &  & 1.74 & 2.40 & & 1.40 & 1.71 & 1.88 \\
    \hline
  \end{tabular}
\end{table}

M2014 constructed the ``band-o-gram'' schematic shown in Figure~\ref{fX}C. The band-o-gram is constructed using only three pieces of information in each cycle: the time of sunspot maxima in each solar hemisphere, and the terminator. The times of hemispheric maxima are determined using the monthly hemispheric sunspot number (e.g. Figure~\ref{fX}), while the terminator times are determined using the sunspot-area threshold described above. The colors of the bands represent the polarities of the underlying magnetic system with red being positive and blue being negative. The demonstrated progression at high latitudes, above 55\degree{}, is defined by the average evolution \citep{2010ApJ...725..658U} of features (7\degree yr$^{-1}$) and was prescribed to be identical for each cycle as there was not enough information to constrain their variability. In the M2014 picture, the time at which the band departs from 55\degree{} in each hemisphere \-- with a poleward and an equatorward branch of the same polarity \-- is set to the time of sunspot maximum for the band at low latitudes in that hemisphere. M2014 then makes a (linear) approximation that the migratory speed from high to low latitudes is set by the hemispheric maximum and terminator at the Equator ($2.75\degree$ yr$^{-1}$). M2014 also identified that time between alternating hemispheric maxima \-- those of the same leading polarity \-- was approximately 21.6 years. This last point permitted the projection of the Cycle 24 bands and the prediction of the onset of the bands that would give rise to Cycle 25 sunspots, both recently confirmed to be following the projected path \citep{2017FrASS...4....4M}. The expectation is that the next terminator will take place late in 2019, or early in 2020, and will trigger the growth of Solar Cycle 25.

\begin{figure}[!ht]
\centering
\includegraphics[width=1.00\linewidth]{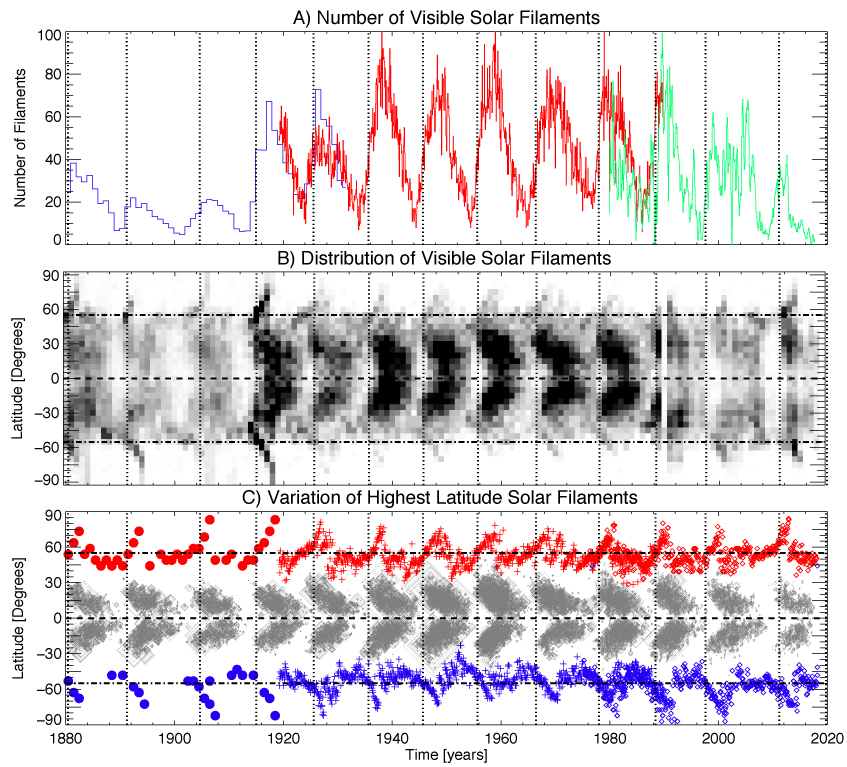}
\caption{The 140 year record of solar filaments as observed in H${\alpha}$ observations from three sites: Arcetri Astrophysical Observatory (AO; 1880\--1929), Meudon Observatory (MO; 1919\--1989), and the Kislovodsk Observatory (KO; 1980\--2018). {\bf Panel A} shows the variation in the total number of filaments observed, for AO the data is provided on an annual basis, MO on a 28 day rotational basis, and KO on a daily basis. {\bf Panel B} shows the composite variation of the annual filament number density as a function of latitude and time ({\it cf.} Figure~\ref{fX}B) in inverted grayscale. {\bf Panel C} contrasts the sunspot butterfly pattern in gray with the modulation of the highest latitude filaments present at each timestep (red:north; blue:south) and each observatory. In each figure we show the locations of the terminators as provided by M2014 as {\it vertical dotted} lines (see above). In Panels B and C we also show the {\it dot\--dashed} lines at 55\degree{} in each hemisphere for reference.}
\label{fY}
\end{figure}

\subsection{Terminators in the Filament Record}
Records of solar features of comparable length to the sunspot observations shown in Figure~\ref{fX} are rare. Figure~\ref{fY} presents a composite record of solar filaments from 1860 to the present. Efforts to detect, catalog, and assess the variation of these features on the solar disk have been many \citep{2014SSRv..186..169C, Gibson2018}. This composite record is constructed from observations taken at the Arcetri Astrophysical Observatory \citep{1933MmSAI...6..479B}, the Meudon Observatory \citep{1975SoPh...44..225H}, and the Kislovodsk Observatory \citep{2016SoPh..291.1115T}.  Note that number density of filaments shown in Figure~\ref{fY}A seldom reaches zero, but it is clearly quasi-periodic like its sunspot counterpart in Figure~\ref{fX}A.

In addition to the correspondence between the terminators and the rapid growth of the sunspot butterfly wings, notice from Figure~\ref{fY} (B and C) that the terminators coincide with a repeated symmetry in filament evolution on this global scale. The enhanced fingers of filament density that extend beyond 55\degree{} latitude belong to an event \citep{1988Natur.333..748W,2014SSRv..186..169C} known as the ``rush to the poles.'' This event occurs once roughly every 11 years and is part of the Sun's polar magnetic-field reversal process \citep{2010LRSP....7....1H}. From Figure~\ref{fY}C we see that the start of the rush to the poles, as visualized in the timeseries of the highest-latitude filament in each hemisphere, appears to be strongly synchronized across the Equator \-- the feature appears to start at identical times in each hemisphere \-- even though hemispheric sunspot activity can often display lags of several years \citep{2010LRSP....7....1H,2013ApJ...765..146M}. Further, it appears that the apparent synchronization at high latitudes is strongly coincident with the terminators of M2014 and the initialization of the butterfly wings. This 140 year record of correlated behavior points to a (strong) coupling between these magnetized systems, and it appears that the terminator plays a critical role, or is a bi-product of this coupling. While we will offer no explanation in this article, the strong repetition of the filament timeseries visible in Figure~\ref{fY}C and the strong bounding of 55\degree{} latitude on filament production should motivate investigations as to the underlying flow patterns at those latitudes, as illustrated in M2014. There would appear to be a fundamental process lurking there. Note that these observations will permit modifications, especially at high latitudes and in early years, to be made to the M2014 band-o-gram, where previously only the sunspot butterfly diagram was available (Figure~\ref{fX}B), but that is beyond the scope of the present work.

 A comparison of Figures~\ref{fX} and~\ref{fY} indicates that the filament record could be used to modify the high-latitude evolution of the band-o-gram: the timing of the rush to the Poles is related to the terminator, not the hemispheric maxima, and that there is considerable variability in the speed of poleward migration about the average. Similarly, the highest latitude filament traces out the path of the polarity-inversion line between the old polar field and the new, and so it could be used to more accurately represent the migratory speed in that part of the system. Such alterations are beyond the scope of the present article.

\subsection{Terminators and Solar Activity Proxy Indices}

\begin{figure}[!ht]
\centering
\includegraphics[width=0.75\linewidth]{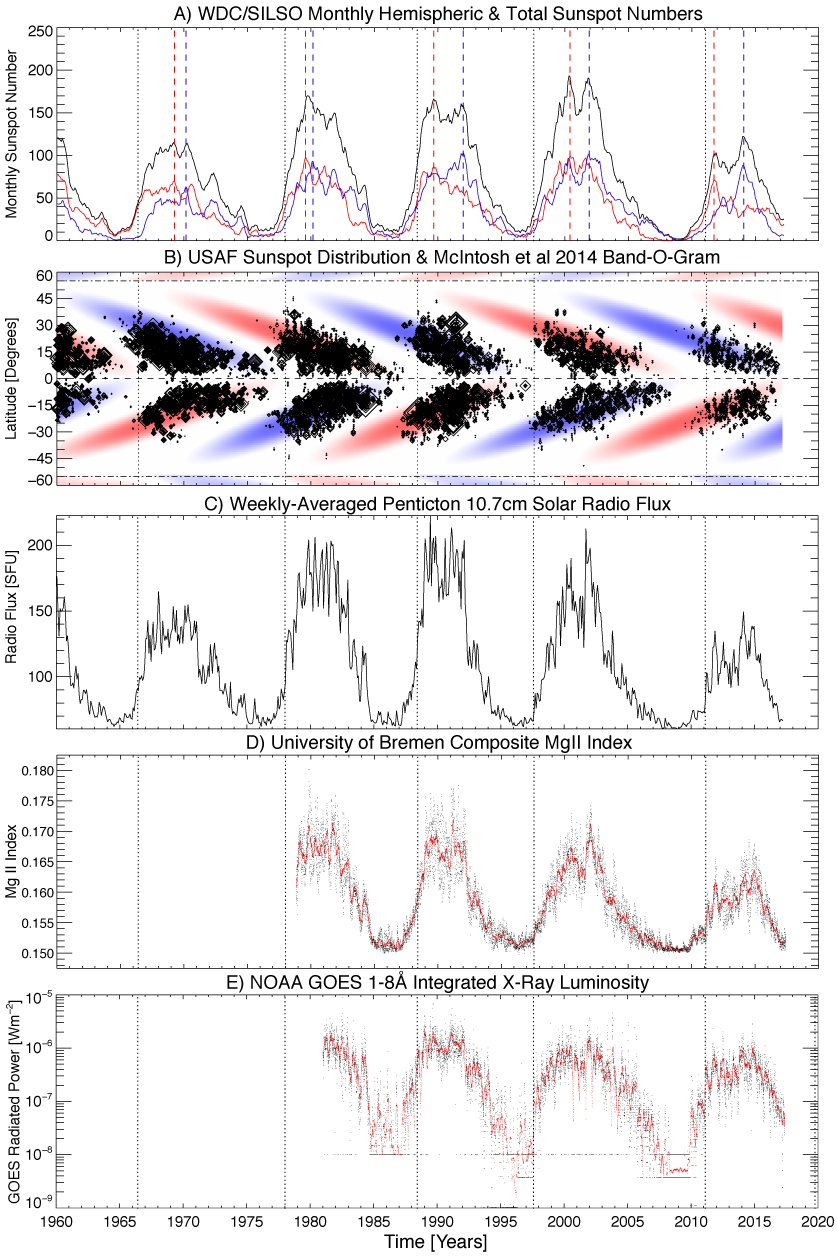}
\caption{Hemispheric sunspot variation, band-o-gram comparisons and the signature that terminators present in Sun-as-a-star proxy measurements over the past 60 years. {\bf Panel A} expands on earlier figures, showing the total ({\it black}) and hemispheric sunspot numbers (red:north; blue:south) with their maxima as correspondingly colored vertical dashed lines. {\bf Panel B} show a comparative overlay of the sunspot butterfly diagram (Figure~\ref{fX}B) and the M2014 band-o-gram (Figure~\ref{fX}C) in this timeframe. {\bf Panel C} shows the weekly average of the Pentincton 10.7\,cm radio flux from the Sun (F10.7). {\bf Panel D} shows the variation of the composite Mg{\sc ii} index provided by the university of Bremen (black) with the 50\-day running average overplotted in red. {\bf Panel E} shows the integrated (1\--8\,\AA) coronal irradiance constructed from the fleet of NOAA GOES spacecraft, again with the 50\--day running average {\it overplotted in red}. In each panel we show the locations of the M2014 terminators as {\it vertical dotted} lines.}
\label{fU}
\end{figure}

Since the dawn of the space age, as a means to evaluate the impact of the Sun on our atmosphere and the near-Earth space environment, the solar research community has made routine Sun-as-a-star activity proxy measurements. Figure~\ref{fU} shows the variation of three such proxies: the 10.7\,cm radio flux (F10.7); the Bremen Mg{\sc ii} index; and the GOES 1-8\AA{} (integrated) x-ray luminosity. Each of the these indices can be compared against the band-o-gram, the butterfly plot and the terminators, as we have above. The chromospheric Mg{\sc ii} index shows a small jump (1--3\,\%) at the terminators, with some being stronger/clearer than others. The two coronal measures, F10.7 and GOES, show jumps in value of order 30 and 50\,\%, respectively. Similarly, the index can have a small up-swing in value before the step function that we might associate with the death of the last cycle at the Equator. These pre-termination upswings in the Sun-as-a-star indices could be caused by pulses of residual equatorial activity \citep[][{\it e.g.}, the 2011 example]{2015NatCo...6E6491M}, and/or by the small spots that occur for several rotations at the same longitude prior to the ramp-up in the butterfly wings \citep[][{\it e.g.}, the 1977 example]{2010LRSP....7....1H}. These abrupt, step-like, changes in solar activity have been previously identified in radiative proxies and have been associated with solar-cycle start-up and solar-cycle prediction \citep{2005MmSAI..76.1034S,2009AdSpR..43..756S}.

\clearpage
\subsection{Contemporary Terminators}
To this point we have dealt with measures that show the presence of terminators, and their apparently abrupt nature. Recalling the statement of M2014 that these events occur in ``around one solar rotation'', can we use contemporary data to investigate how quickly the transition between solar cycles might be?

\begin{figure}[!ht]
\centering
\includegraphics[width=1.00\linewidth]{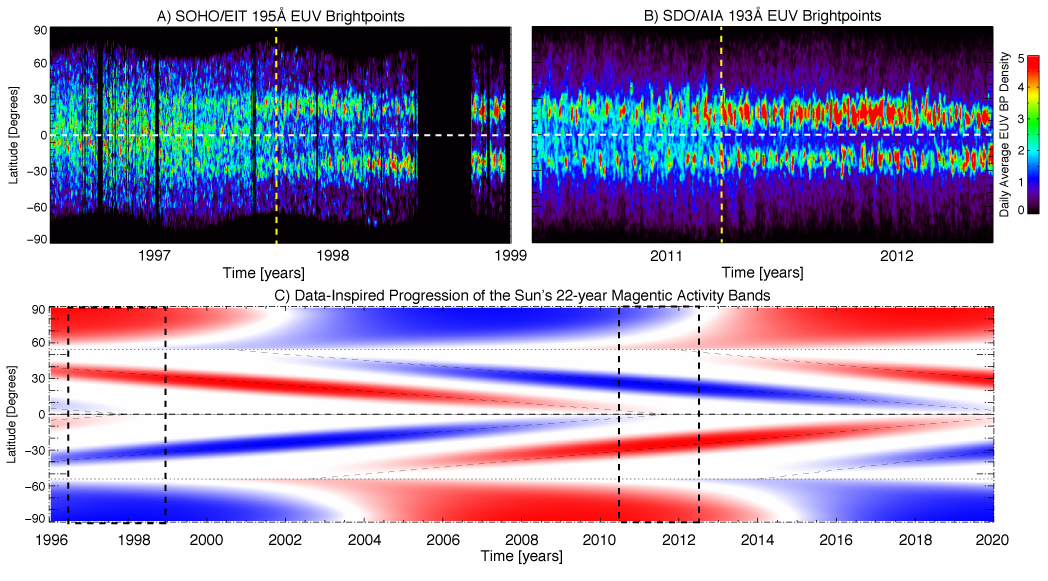}
\caption{Visualizing the termination of magnetic activity bands. Latitude-time variation of EUV Bright Points (BPs) for \soho{}/EIT 195\AA{} through the Solar Cycle 22/23 minimum (1997\,\--\,1999; {\bf Panel A}) and the Solar Cycle 23/24 minimum (2010\,\--\,2012; {\bf Panel B}) for \sdo{}/AIA 193\,\AA{}. The rapid decrease, or ``termination,'' of the activity related to EUV BPs at the Equator is clear. The apparent termination of the bands belonging to the 22-year solar magnetic activity cycle are marked, by eye, as August 1997 and February 2011 with vertical dashed yellow lines. Panel C shows the band-o-gram of Figure~\ref{fX}, reduced to the \soho{}/\stereo{}/\sdo{} epoch. The {\it red and blue colored migrating bands} illustrate a positive and negative magnetic polarities, respectively. The {\it boxed regions} in {\bf Panel C} illustrate the time-frame studied in Panels A and B and below.}
\label{f1}
\end{figure}

Figure~\ref{f1} shows a six-month span around the 1997 (Panel A) and 2011 (Panel B) termination events to highlight the abrupt change in BP density occurring at the end of Solar Cycles 22 and 23, respectively. As above, the band-o-gram portion in Panel C, illustrates the interaction and progression of the overlapping, polarized magnetic bands with latitude in the same timeframes. We see that despite the differing spatial resolutions of EIT and AIA and the lower cadence of images used (four per day for EIT, twenty-four per day for AIA), the changes to the system seem to be grossly reproduced in these events that are approximately fourteen years apart. Note that the BP density plots shown in this article illustrate the latitudinal distribution of BPs over the 5\degree{} wide band centered on the Sun's central meridian relative to each spacecraft.

\begin{figure}[!ht]
\centering
\includegraphics[width=1.00\linewidth]{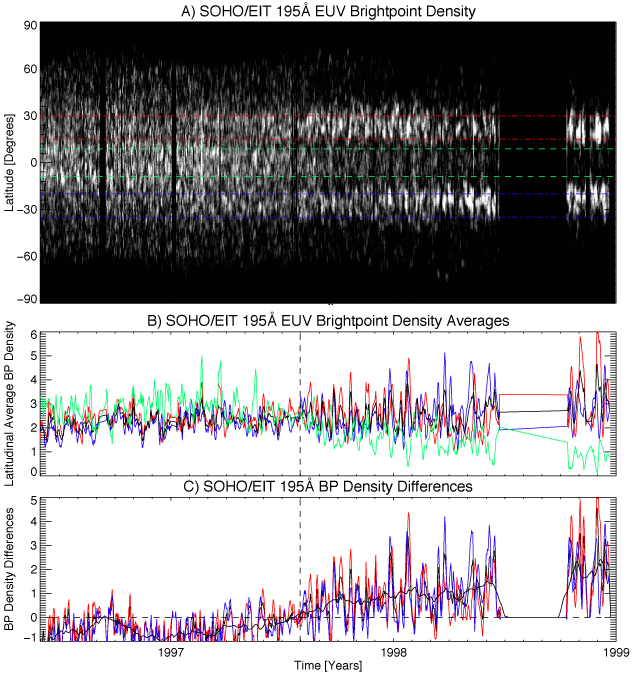}
\caption{The 1997 transition between Solar Cycles 22 and 23 from the perspective of \soho{}. Comparing {\bf Panel~A} with the corresponding Panel of Figure~\ref{f1} we show the latitude\--time variation of EUV BPs for \soho{}/EIT 195\,\AA{} through the Solar Cycle 22/23 minimum (1997--1999). On this plot three regions are identified, one equatorial (bounded at $\pm7.5\degree$ latitude by {\it dashed green lines}), one on the northern activity band (15\degree{} wide, centered on 22.5\degree{} latitude, bounded by {\it dashed red lines}), and another on the southern activity band (15\degree{} wide, centered on $-27.5\degree{}$ latitude, bounded by {\it dashed blue lines}). {\bf Panel~B} shows the corresponding latitudinally-averaged BP density in {\it green, red, and blue} while the {\it black line} illustrates the average BP density of two mid-latitude bands. {\bf Panel~C} shows the difference between the equatorial and northern BP density ({\it red}), the equatorial and southern BP density ({it blue}) and the equatorial and hemispherically averaged BP density (black). Also shown is a 28-day running average of the equatorial and hemispherically averaged BP density: the {\it smoother black line}. The {\it dashed vertical lines} in each Panel are drawn at 1 August, 1997.}
\label{f2}
\end{figure}

Figure~\ref{f2} shows the 1997 terminator in more detail, noting that the BP-density plots shown are saturated to bring out weak variations. In Figure~\ref{f2}A we delineate the low-latitude [$\pm~7.5\degree$] by dashed green lines from the northern (15\degree{} wide and centered on 22.5\degree{}) and southern (15\degree{} wide and centered on $-27.5\degree{}$) hemispheric mid-latitude bands using red and blue dashed lines, respectively. The central latitude and width of the activity bands ($15\degree$) in each case studied herein have been established by Gaussian fits to the latitudinal BP density distributions as illustrated in Figure~3 of M2014. The equatorial band width was chosen to match those of the mid-latitudes. The average BP density in each region is shown in Figure~\ref{f2}B by traces of the same color and introducing an average of the mid-latitude activities with a black line. We note that the Solar Cycle 23 activity bands are offset slightly in latitude due to the two-year hemispheric phase lag in activity \citep{2013ApJ...765..146M}. Since we wish to study the relative evolution of the mid-latitude activity bands compared to the equatorial region, Figure~\ref{f2}C shows the difference in the BP density between the equatorial and hemispheric bands as blue and red curves while the difference between the equatorial and hemispheric average is black. We note that the dashed vertical line is midnight (universal time) on 1 August, 1997 and that the tick marks on the horizontal (time) axis are one terrestrial month apart. That point in time sees a 100\,\% growth in BP density on the mid-latitude bands and a 100\,\% decline in BP density at low latitudes within only a small number of solar-rotations. Note also that following the termination point, there is a significant increase in the variability of the signal in the activity bands in both hemispheres \-- increasing the RMS of the BP density by an average of one BP per degree per day. In other words, the activity bands at the Equator significantly decreased in activity while those at mid-latitudes saw a significant increase at approximately the same time.

The EIT observations support the M2014 picture \-- one where the oppositely polarized magnetic activity bands of the 22-year magnetic activity cycle meet and cancel at the Solar Equator. It would appear that the occurrence of this ``event'' at the Equator is abrupt with the large change in the mid-latitude BP density occurring in close conjunction. Rapid growth of activity occurs on those mid-latitude bands \-- presumably due to a rapid increase in the rate of magnetic flux emergence therein, following the termination. 
Based on the EIT observations and analysis, the switch\-over between low- and mid-latitudes occurred on a timescale less than a rotational period, or 28 days, but we cannot be much more precise \-- better signal, and/or multi-viewpoint observations are required.

\begin{figure}[!ht]
\centering
\includegraphics[width=1.00\linewidth]{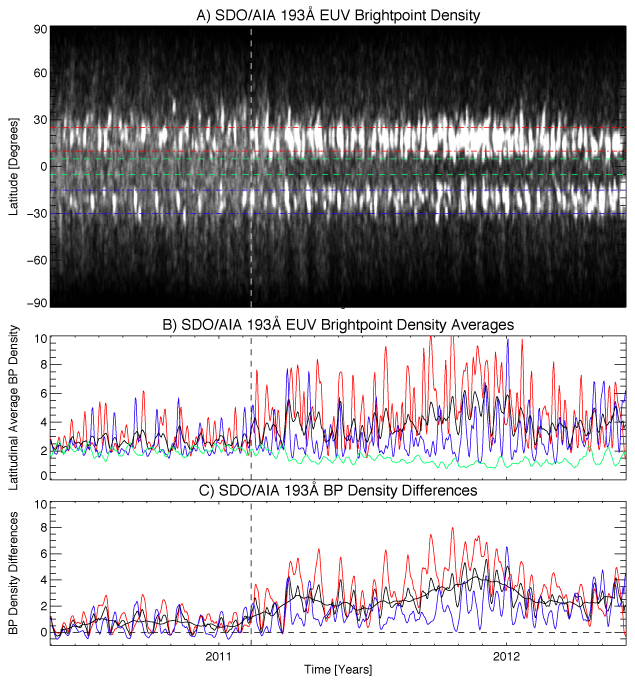}
\caption{The 2011 transition between Solar Cycles 23 and 24 from the perspective of \sdo{}. Comparing {\bf Panel~A} with Panel B of Figure~\ref{f1} we show the latitude--time variation of EUV BPs for \sdo{}/AIA 193\,\AA{} through the Solar Cycle 23/24 minimum (2010\,\--\,2012). Three regions are identified: one equatorial (bounded at $\pm5\degree$ latitude by green dashed lines), one on the northern activity band (15\degree{} wide, centered on 17.5\degree{} latitude, bounded by {\it dashed red lines}), and another on the southern activity band (15\degree{} wide, centered on $-22.5\degree{}$ latitude, bounded by {\it dashed blue lines}). {\bf Panel~B} shows the corresponding latitudinally--averaged BP density in {\it green, red, and blue} while the {\it black line} illustrates the average BP density of two mid-latitude bands. {\bf Panel~C} shows the difference between the equatorial and northern BP density ({\it red}), the equatorial and southern BP density ({\it blue}) and the equatorial and hemispherically averaged BP density ({\it black}). Also shown is a 28-day running average of the equatorial and hemispherically averaged BP density \-- the smoother black line. The vertical dashed lines in each Panel are drawn at 11 February, 2011.}
\label{f4}
\end{figure}

\begin{figure}[!ht]
\centering
\includegraphics[width=1.00\linewidth]{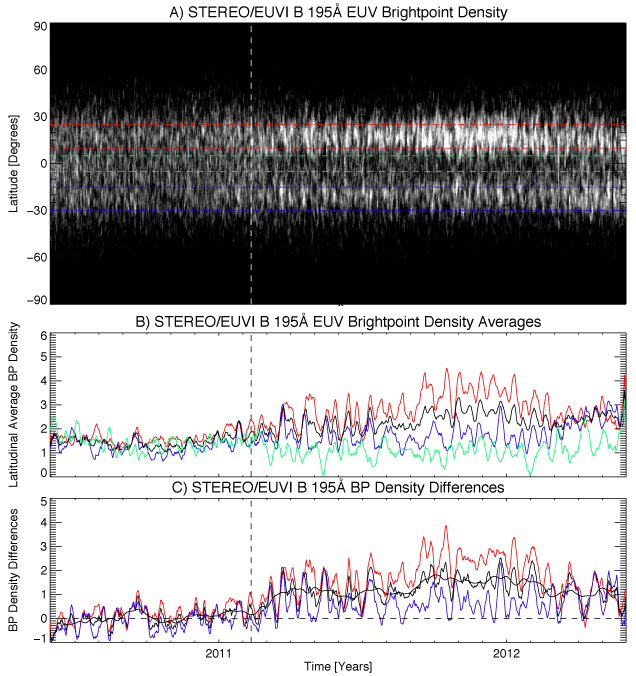}
\caption{The 2011 transition between Solar Cycles 23 and 24 from the perspective of \stereo{}-B. We show the latitude-time variation of EUV BPs for \stereo{}-B/EUVI 195\,\AA{} through the Solar Cycle 23/24 minimum (2010\--2012). On this plot three regions are identified, one equatorial (bounded at $\pm5\degree$ latitude by {\it dashed green line}), one on the northern activity band (15\degree{} wide, centered on 17.5\degree{} latitude, bounded by {\it dashed red lines}), and another on the southern activity band (15\degree{} wide, centered on $-22.5\degree{}$ latitude, bounded by {\it dashed blue lines}). {\bf Panel~B} shows the corresponding latitudinally-averaged BP density in green, red and blue while the black line illustrates the average BP density of two mid-latitude bands. {\bf Panel~C} shows the difference in between the equatorial and northern BP density ({\it red}), the equatorial and southern BP density ({\it blue}) and the equatorial and hemispherically averaged BP density ({\it black}). Also shown is a 28-day running average of the equatorial and hemispherically averaged BP density \-- the {\it smoother black line}. The {\it dashed vertical lines} in each Panel are drawn at 11 February, 2011.}
\label{f3}
\end{figure}

\begin{figure}[!ht]
\centering
\includegraphics[width=1.00\linewidth]{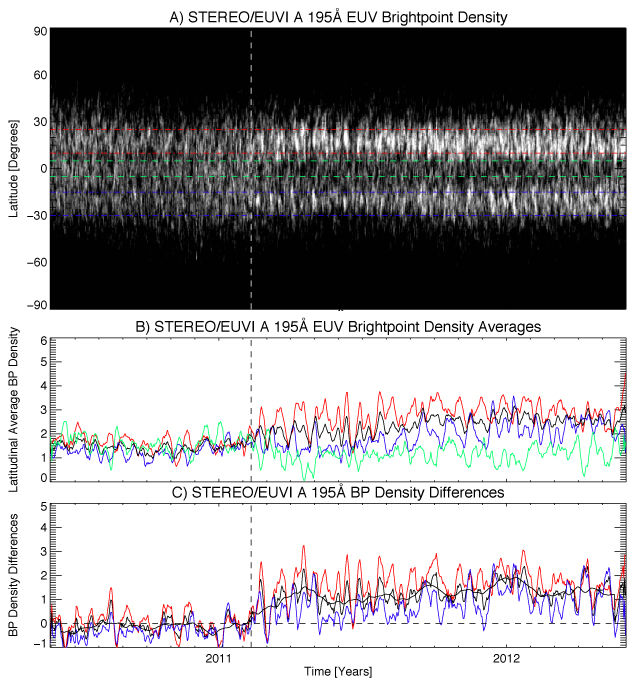}
\caption{The 2011 transition between Solar Cycles 23 and 24 from the perspective of \stereo{}-A. We show the latitude-time variation of EUV BPs for \stereo{}-A/EUVI 195\,\AA{} through the Solar Cycle 23/24 minimum (2010\--2012). On this plot three regions are identified, one equatorial (bounded at $\pm5\degree$ latitude by {\it dashed green line}), one on the northern activity band (15\degree{} wide, centered on 17.5\degree{} latitude, bounded by {\it dashed red lines}), and another on the southern activity band (15\degree{} wide, centered on $-22.5\degree{}$ latitude, bounded by {\it dashed blue lines}). {\bf Panel~B} shows the corresponding latitudinally-averaged BP density in green, red and blue while the black line illustrates the average BP density of two mid-latitude bands. {\bf Panel~C} shows the difference in between the equatorial and northern BP density ({\it red}), the equatorial and southern BP density ({\it blue}) and the equatorial and hemispherically averaged BP density ({\it black}). Also shown is a 28-day running average of the equatorial and hemispherically averaged BP density \-- the {\it smoother black line}. The {\it dashed vertical lines} in each Panel are drawn at 11 February, 2011.}
\label{f5}
\end{figure}

Figures~\ref{f4}, ~\ref{f3}, and~\ref{f5} show the equivalent plots of Figure~\ref{f2} for the termination event of 2011 from the \sdo{} and the two \stereo{} spacecraft, respectively. During this time period the observations from the three spacecraft simultaneously covered every solar longitude, marking the first time that such observations were available. Ordered in advancing heliographic longitude, where Figure~\ref{f3} explores the latitudinal BP density observed from \stereo{}-B \-- behind the Sun-Earth line, Figure~\ref{f4} from \sdo{} on the Sun-Earth line ({\it cf.} Figure~1), and Figure~\ref{f5} shows that from \stereo{}-A \-- ahead of the the Sun-Earth line. In each figure the vertical dashed lines mark a reference date of midnight (universal time) on 11 February, 2011 and, again, the tick marks on the horizontal (time) axis are one terrestrial month apart. Also, given the longitudinal separation of the observations, these three figures represent BP evolution centered on different longitudes \-- a feature we will take advantage of below. For this timeframe we delineate the low latitude region ($\pm~5\degree$) by dashed green lines from the northern (15\degree{} wide and centered on 17.5\degree{}) and southern (15\degree{} wide and centered on $-22.5\degree{}$) hemispheric mid-latitude bands using dashed red and blue lines, respectively. As above, the band thickness ($15\degree$) is established by Gaussian fits to the latitudinal BP density distributions from M2014 specific to Solar Cycle 24. The closer proximity of the mid-latitude active bands to the Equator means that limit the equatorial region to only 10\degree{} wide to minimize the inclusion of that signal in the analysis (see below). The information in the three figures quantitatively behave in a similar fashion to that of Figure~\ref{f2}; approximate order of magnitude decrease at the Equator and a step-like, rapid, increase at mid-latitudes occurring on the timescale of a solar rotation. In each case we also note the relative post-termination growth in the southern hemisphere is about half of that in the North. We believe that this difference is tied to the two-year hemispheric activity asymmetry exhibited during this epoch, mentioned earlier \citep{2013ApJ...765..146M}.

\begin{figure}[!ht]
\centering
\includegraphics[width=0.65\linewidth]{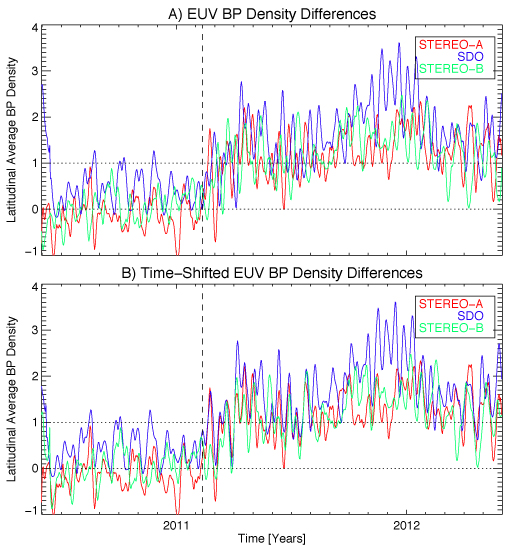}
\caption{Comparing the difference in equatorial and mid-latitude average BP densities from three longitudinal vantage points. {\bf Panel A} shows the equatorial-to-mid-latitude differences in BP density from {\em STEREO-A}/EUVI, \sdo/AIA, and {\em STEREO-B}/EUVI are shown as {\it red, blue, and green lines}, respectively ({\it cf.} Figures~\ref{f4},~\ref{f3}, and~\ref{f5}). {\em Panel B} shows the same BP density differences shifted by the maximum lag of their cross-correlation (Figure~\ref{f7}). For reference in each panel we show a {\it dashed vertical line} at 11 February, 2011 and {\it dotted lines} at differences of 0 and 1 to illustrate the jump in values.}
\label{f6}
\end{figure}

\begin{figure}[!ht]
\centering
\includegraphics[width=0.50\linewidth]{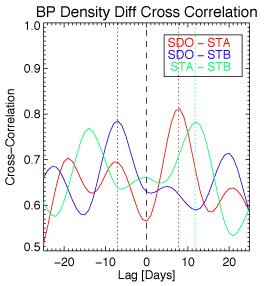}
\caption{BP density difference cross-correlation functions. We show the cross-correlation functions of the equatorial-to-mid-latitude differences in BP density between \sdo{}/AIA and {\em STEREO-A}/EUVI ({\it red}), \sdo{}/AIA and {\em STEREO-B}/EUVI ({\it blue}) and the twin \stereo{} instruments ({\it green}). In each case we demonstrate the peak cross-correlation lag by {\it dotted vertical lines}. For reference the {\it dashed vertical line} shows a lag of zero days.}
\label{f7}
\end{figure}

\begin{figure}[!ht]
\centering
\includegraphics[width=1.00\linewidth]{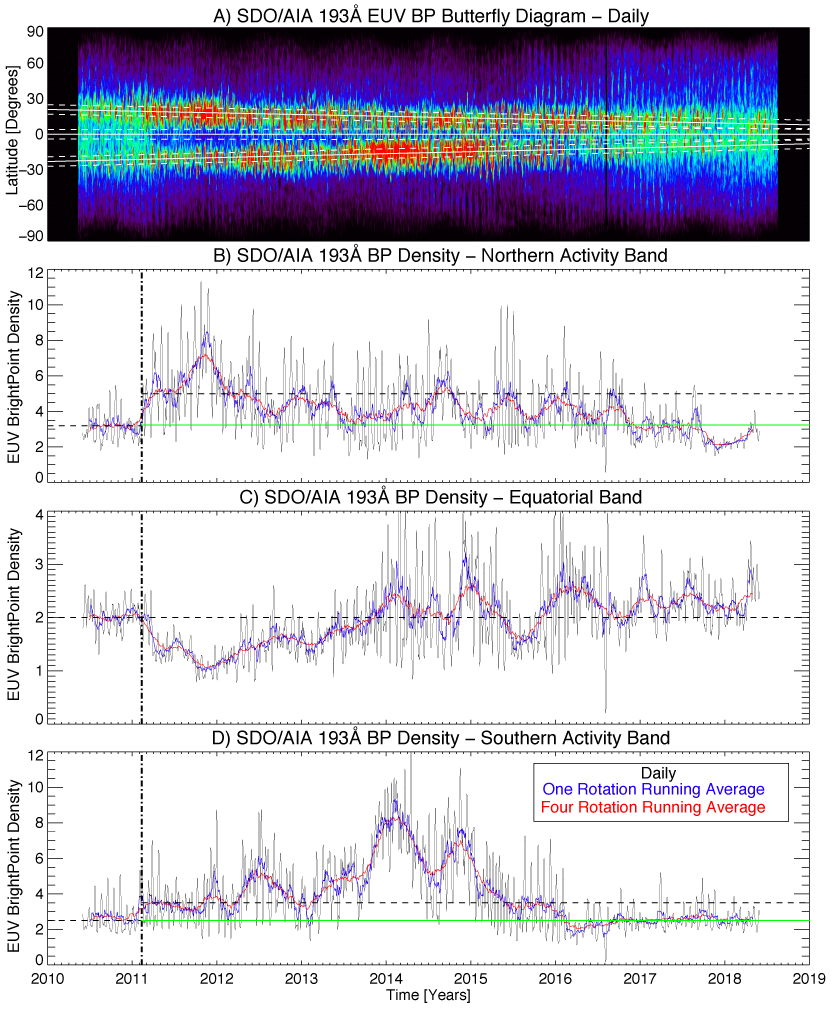}
\caption{Comparing the EUV BP density butterfly plot and traces of activity over the \sdo{} mission to date. From top to bottom, the BP density of Figure~\ref{f4} extended to the eight year mission with the best-fit activity band centroids an equatorial band shown as solid lines with only a 10\degree{} wide region around their center. {\bf Panel B} through {\bf Panel D} show the variability along each band, north, Equator and south respectively. Each plot shows the BP density averaged over the band ({\it black}), that averaged over one rotation in {\it blue} and over four solar rotations in {\it red}. For reference the averages of the six months before and after the 2011 termination point are shown as {\it dashed lines}. Panels B and D show the ``before'' line continued after the termination event as a {\it dashed green line}. In these plots the step-like variation immediately following the termination event is clear. Similarly, it is clear that while the BP densities exhibit smooth variability following the terminator, they do not return to their pre-event values for many years. The equatorial region recovers first, but that is likely the result of the encroachment of the mid-latitude bands in to the equatorial range of latitudes.}
\label{fA}
\end{figure}

Figure~\ref{f6} compares the Equator to mid-latitude average BP density differences in these three cases. Comparing the three curves in Panel A we can see that the signals around February~11, 2011 are longitudinally offset from one another. Cross-correlation of the timeseries (Figure~\ref{f7}) indicate that the {\em STEREO-A} signal leads that of AIA by 8 days and that of {\em STEREO-B} by 15 days. Panel B shows the BP density differences time-shifted by the appropriate lag-times and we see a strong correspondence in the timeseries, especially around February 11, 2011. Note that, towards the end of the timeseries shown, those lag-time values become increasingly inaccurate due to the continued drifting of the \stereo{} spacecraft relative to the Sun-Earth line which slowly increases the longitudinal separation of the signals. The temporal offset between the three spacecraft would appear to indicate that the cycle termination event is {\em strongly} longitudinal and rapid in nature--possibly occurring in a fraction of a rotation. Following the termination, activity at mid-latitudes, {\em and following longitudes}, is rapidly elevated to a new level. This new plateau of activity is shown in Figure~\ref{fA} from the \sdo{} perspective--the Sun does not appear to drop below that level for many years.

\begin{figure}[!ht]
\centering
\includegraphics[width=1.00\linewidth]{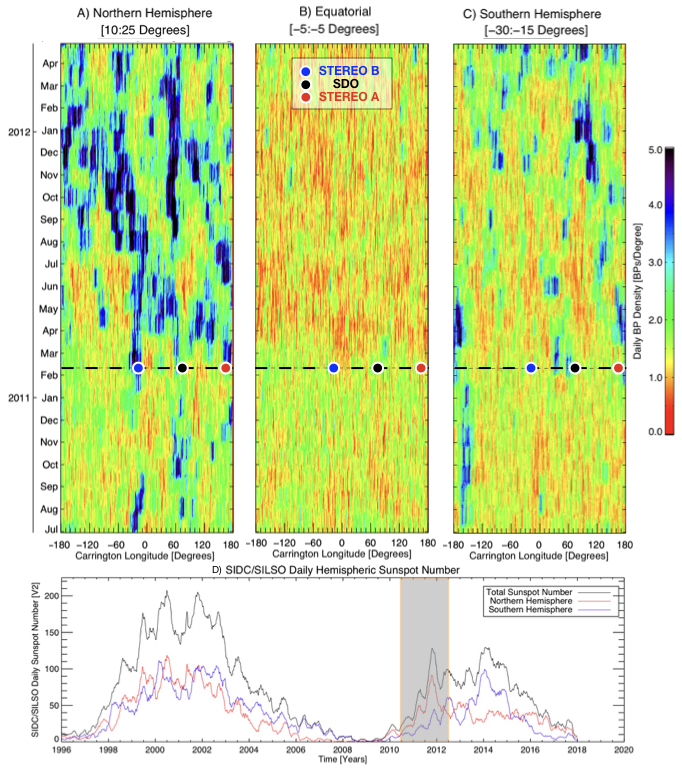}
\caption{The longitudinal evolution of the BP density around the 2011 terminator for the northern activity band ({\bf A}), equatorial region ({\bf Panel B}), and southern activity band ({\bf Panel C}) as observed by \sdo{} and the twin \stereo{} spacecraft. Compare these ``Hovm\"{o}ller'' diagrams with Figures~\ref{f4}, \ref{f3}, \ref{f5}, Figure 9.3 of Wilson \citep{1994ssac.book.....W}, and Figure 3 of \cite{2017NatAs...1E..86M} who first identified magnetized Rossby waves in the solar interior. The BP density is constructed using a 28-day running average of instantaneous BP density observed by the three spacecraft \-- as introduced in the aforementioned article. The {\it dashed horizontal lines} on each panel designate 11 February, 2011 and the {\it colored dots} on those lines indicate the longitude of the central meridian as seen by each spacecraft on that date. {\bf Panel D} shows running 50-day averages of the daily total and hemispheric sunspot numbers. The {\it gray shaded box} outlines the timeframe shown in the upper panels. It should be used by the reader to relate longitudinal surges of solar activity on the activity bands and the gross reduction of activity in the equatorial region that follow the termination. }
\label{fR}
\end{figure}

\begin{figure}[!ht]
\centering
\includegraphics[width=1.00\linewidth]{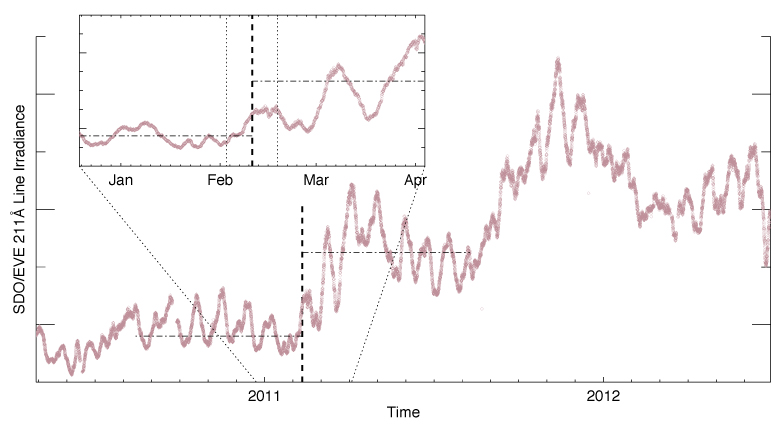}
\caption{The evolution of coronal emission seen from Sun-as-a-Star \sdo{}/EVE spectral irradiance measurements. In the main body of the figure we show the variation of the Fe~{\sc xiv} 211\,\AA{} coronal emission line as a function of time. The vertical dashed line indicates 11 February, 2011 at 00:00UT. The data are normalized such that a large tick mark indicates 100\,\% change after that date. The horizontal lines indicate average activity levels over the three months before and after 11 February, 2011. The Inset panel shows the same radiative variability over two solar rotations before and after 11 February, 2011. Again, the {\it dashed vertical line} indicates 11 February, 2011 at 00:00UT and the {\it dotted vertical lines} are placed 8 days before and after. One can readily compare the shape of this radiative profile with the EUV BP density plots in Figure~\ref{f6}.}
\label{f9}
\end{figure}

\begin{figure}[!ht]
\centering
\includegraphics[width=1.00\linewidth]{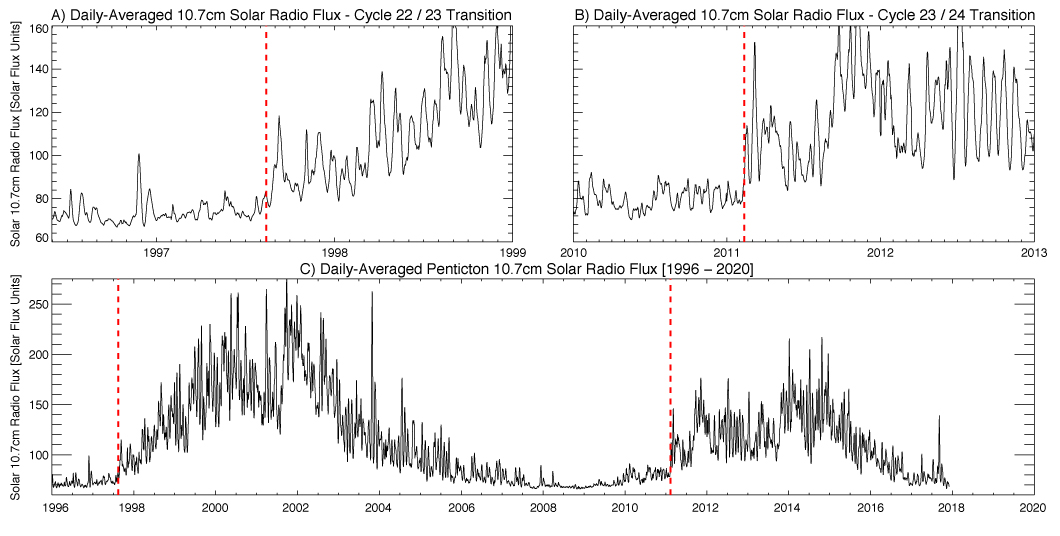}
\caption{The evolution of the daily 10.7cm radio flux over the epoch studied. The 10.7cm radio flux is a standard Sun-as-a-Star proxy for coronal activity  \citep{2013SpWea..11..394T}, where we compare radio activity around the Solar Cycle 22/23 minimum (1997 \-- 1999; {\bf Panel A}) the Solar Cycle 23/24 minimum (2010 \-- 2012: {\bf Panel B}) and for the entire 22-year period from 1996 to 2018 as in Figure~\ref{f1} in {\bf Panel C}. In each panel the {\it dashed red vertical lines} indicate the times of the cycle 22 and 23 band terminations. Note the close correspondence between {\bf Panel B} and the \sdo{}/EVE irradiance measurement shown in Figure~\ref{f9}.}
\label{f10}
\end{figure}

\begin{figure}[!ht]
\centering
\includegraphics[width=1.00\linewidth]{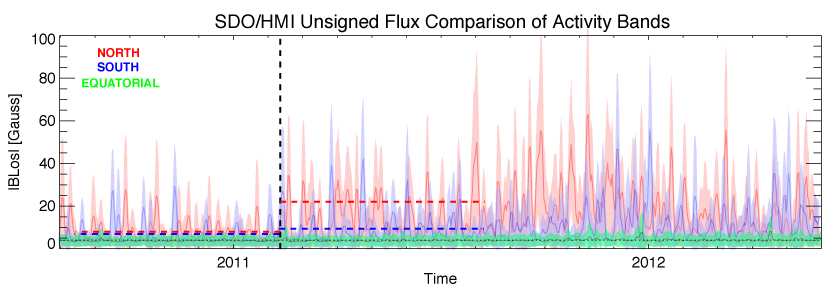}
\caption{Comparing the absolute value of the average solar magnetic flux in the activity bands of Figures~\ref{f4} and~\ref{f5} used above. The {\it red}, {\it green}, and {\it blue shaded regions} respectively show the mean value and one standard deviation of variability across the band in each case. The {\it dashed vertical line} indicates 11 February, 2011 at 00:00UT. The {\it horizontal lines} indicate respective average activity levels over the three months before and after 11 February, 2011.}
\label{f8}
\end{figure}

Figures~\ref{f6} and~\ref{fA} can be compared with Figure~\ref{fR} that shows \sdo{}/\stereo{} BP Hovm\"{o}ller diagrams \citep[longitude\--time plot at fixed latitudes, e.g.][]{2017NatAs...1E..86M}  of the same activity and equatorial regions as explored in Figures~\ref{f4}, \ref{f3}, and \ref{f5}. Note that activity is reduced rapidly at a narrow range of longitudes at the Equator before rapidly spreading across all longitudes (within one rotation). In the same epoch several longitudes are enhanced on the activity belts at the terminator with more longitudes growing subsequently, and in the North notably within two or three rotations. This step-like longitudinal growth of activity is supported for the 2011 event by recent investigations of the Sun-as-a-star irradiance at many wavelengths ({\it e.g.}, Figures~\ref{f9} and \ref{f10}) and by a host of other resolved observations from space  \citep{0004-637X-844-2-163, Morgane1602056}, including the unsigned magnetic field in the activity bands ({\it e.g.}, Figure~\ref{f8}).

\clearpage

\section{Discussion}
In the preceding sections we have shown multiple examples of what appears to be significant times in the progression of solar-activity; termination points, or ``terminators,'' going back nearly 140 years in the observational record. First identified by M2014, these events marked the transition from the remnants of the last Solar Cycle to the beginning of the next one as the times of significant sunspot occurrence after solar minimum. Those events appeared to take place on a timescale close to one solar rotation. Comparing the sunspot progression with the record of the highest-latitude H${\alpha}$ filaments over the same epoch we see that the terminators not only coincided with the rapid growth of Maunder's butterfly wings at around $35\degree$ latitude, but they also coincide with the initialization of the ``rush to the poles'' in both solar hemispheres in all examples observed \-- some thirteen Solar Cycles. We note again the synchronization of the latter even though, in most cases, sunspot activity is asymmetric across the Equator by as much as a few years. 

Considering indices of solar activity that are available from the 1960s, we see that the terminators demonstrate a very rapid jump in activity. The upward step in the proxy appears to grow with the apparent temperature of the region where the proxy is formed--lower in the chromosphere and significantly higher in the corona, as also pointed out in the community \citep{2005MmSAI..76.1034S,2009AdSpR..43..756S,0004-637X-844-2-163, Morgane1602056}. It is beyond the scope of this article to explore in depth, but it would be a fruitful exercise to collate the various proxy measurements and apply statistical methods \citep{Basseville:1993:DAC:151741} such as  ``step detection'' to more precisely identify and catalog the termination events and investigate their multi-spectral magnitude.

Another future avenue of exploration would be to search for the termination signature in helioseismic data.  \citet{2016ApJ...828...41S} identified a latitudinal dependence in frequency shifts obtained from GONG data for Cycles 23 and 24, with the region below 15\degree{} latitude showing a delayed, but steeper rise in this sub-surface activity metric, thereby indicating an equatorward propagation of the activity that was shown to correspond well to surface sunspots.  The cycle 22 and 23 termination events of Table \ref{T1} roughly correspond to the minimum of symmeterized low-latitude ($|\theta| < 15\degree$) frequency shifts found in that work, possibly signalling the disappearance of sub-surface fields at those latitudes. However, the time resolution is poor due to the significant (1 year) averaging.  A finer-grained correspondence might be obtained using \sdo{}/HMI Doppler data and less aggressive averaging.

We have presented data from four contemporary spacecraft spanning the timeframes that include the termination of Solar Cycles 22 (in 1997) and 23 (in 2011) through the observations of BPs. The 2011 event, in particular, illustrates the dramatic change in solar flux emergence between the low- and mid-latitudes associated with a terminator. Based on the analysis of the \sdo{} and \stereo{} BP densities we deduce that the timescale for the terminator could be less than one solar rotation. The joint analysis, covering all solar longitudes, indicates that the terminator is a very strongly longitude-dependent event. Observed by each spacecraft, the transition of activity occurs very rapidly at a few longitudes before spreading to more longitudes over the next couple of rotations.

\subsection{Potential Mechanisms}

Clearly, given the broad range of signatures observed, these termination events mark abrupt changes in the Sun's magnetic variability. In the following paragraphs we will discuss a number of possible mechanisms for how such abrupt changes in solar magnetic output could occur. 

\subsubsection{Magnetic Teleconnection}

M2014 inferred that the modulation of the Solar Cycle occurred via a process that they dubbed ``magnetic teleconnection.'' M2014 defined this term as an analog to the term ``teleconnection,'' which refers to climate anomalies being related to each other at large distances on the globe \citep{doi:10.1029/97JC01444}. In short, magnetic teleconnection was proposed to describe the magnetic interactions inferred to take place between the oppositely polarized, overlapping magnetic bands of the 22-year long solar magnetic activity cycle within a solar hemisphere and across the solar Equator (see, {\it e.g.}, Figure~8 of M2014). In that picture, termination events signal the end of one magnetic (and sunspot) cycle at the solar Equator and the start of the next solar cycle at mid-latitudes \-- acknowledging that the bands of the magnetic cycle have been present for some considerable time before the termination. Now, we have seen that they likely also relate to the start of the rush to the poles, and the polar reversal process, at high latitudes. If so, the speed over which these repeated transitions occur could provide critical physical insight into coupling of the activity bands in the solar interior, if that occurs at all.

``Magnetic Teleconnection'' is a phenomenological process to describe how the large-scale magnetic systems of the solar interior of the Sun's 22-year magnetic-activity cycle might interact to produce the 11-year modulation of sunspots and the butterfly diagram (see Figure~\ref{fX}). The interested reader can refer to Figure~8 of M2014, which pictorially describes the process, but also to Figures.~\ref{fX},~\ref{fU} and~\ref{f1}. Our use of the term draws from the meteorological equivalent--teleconnection \citep{Barnston1987} \-- where atmospheric low-frequency variability (such as planetary waves) over timescales of several weeks to several years, are temporally correlated between physical locations that are geographically separated. The North Atlantic Oscillation (NAO) and the El Nino Southern Oscillation (ENSO) are examples of teleconnection phenomena. It is accepted that gravity waves play the role of moderator between the various circulatory patterns and planetary waves.

\subsubsection{Estimating Properties of Teleconnection Processes}
In a magnetized convecting plasma ``information'' can be carried by Alfv\'{e}n waves \citep{1942Natur.150..405A}. The relationship between the measured Alfv\'{e}n speed [$V_{A}$], plasma density, and magnetic-field strength [$V_{A} = B / \sqrt{4\pi\rho}$] gives us the ability to estimate the mean magnetic field strengths in the domain the information that travels through from the transit-time estimate determined above. An Alfv\'{e}n transit time of a few days, using eight days as a possible lower bound, with a transit distance of $\pm25\degree$ ($\approx$219~Mm) to the mid-latitude of the new activity bands, would require the presence of magnetic fields (which can be in fibril form) with a poloidal field strength of about 50\,kG. Assuming an averaged tilt angle of 7\degree{} for the fields, this would imply a total field strength of about 400\,kG, a magnitude supported by estimated rise times of strong magnetic flux elements at mid-latitudes \citep{1983A&A...122..241M}. This is considerably larger than the 40~kG to 100~kG field strength range anticipated in that region, based on calculations of rising thin flux tubes that can produce tilt angles of solar active regions that are consistent with observations. Note that, for reference, state-of-the-art convective dynamo models \citep{2014ApJ...789...35F} are able to produce mean fields of the order of 7~kG although localized regions can reach as much $\approx$30\,kG. 

Further, if Alfv\'{e}n waves are responsible for carrying the information from the termination of one cycle to the beginning of a new cycle at mid-latitudes in a matter of a few days, then it would appear that the Sun's dynamo may be operating in a regime where the magnetic fields present in the Sun's convection zone are dynamically important. That is, the magnetic-field strength is in a state of ``super-equipartition'' relative to convection. While there have been limited theoretical considerations \citep{1994A&A...289..949F}, there are no numerical convective dynamo models operating in that regime despite indications that considerably increasing simulation resolution \citep{2016Sci...351.1427H} leads to a trend of developing small scale magnetic fields that become significantly super-equipartition compared to the convective flows.

In a similar vein, we estimate the approximate timescale for giant-cell convection to carry the information about the cycle termination from Equator to mid-latitudes. Based on estimates of the turbulent diffusion time scale at a giant cell scale [$L_{GC}$] of about 400\,Mm, and a typical kinetic energy density of 10$^6$ erg cm$^-3$, gives a velocity magnitude [$v_{GC}$] of about 34 $\mbox{ m s}^{-1}$. In this case, the $\approx$25\degree{} distance  (D=219 Mm) via turbulent diffusion would take about $D^2/(v_{GC} L_{GC})$, or around 40 days, still longer than that observed, but not by much.

\begin{figure}[!ht]
\centering
\includegraphics[width=1.00\linewidth]{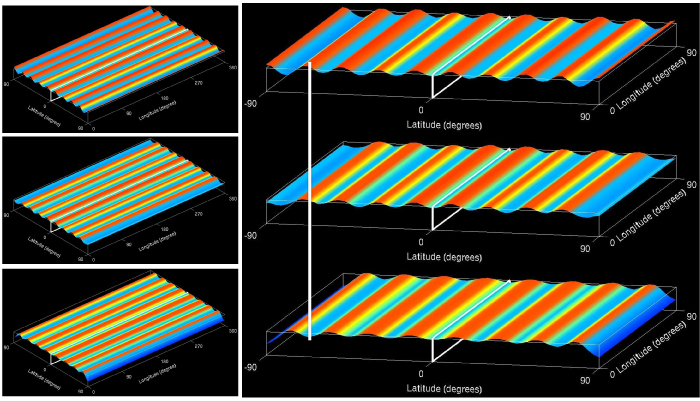}
\caption{Gravity waves produced in a shallow-water \citep{Matsuno1966, 1982bsv..book.....P} tachocline model are shown in two perspective views (see \citet{2019NatSR...9.2035D} for full details). This illustrative case is for the stratification parameter ($\delta$=0.001), characteristic of overshoot tachocline. {\bf Left panels} represent gravity waves viewed along longitude, and {\bf right panels} along latitude. {\it Red-orange-yellow} denotes the crest of the wave ({\it red} denoting the highest part and {\it yellow} the lowest part of the crest); {\it indigo-blue-skyblue} denotes the trough. {\it Green} is neutral. Time goes from top to the bottom. Tachocline gravity waves propagate from the Equator to the pole. The {\it vertical white line} is overlaid on the {\bf right panels} to show the propagation of the wave: in the {\bf topmost} panel the line coincides with the top part of a crest and in the {\bf bottommost} panel the line coincides with the trough of the next wave. The time difference between the {\bf topmost} and {\bf bottommost} frames is about (0.5 \-- 1) days in this example. Therefore it should take about 2.5 to 5 days for a gravity-wave signal to propagate from the Equator to mid-latitudes. Note that in these tachocline-dynamics simulations, longitudinally propagating non-axisymmetric modes appear together with gravity waves propagating in all directions. For simplicity, all non-axisymmetric modes (Rossby and gravity waves) have been removed, so that the latitudinally propagating gravity waves are clearly visible.}
\label{f12}
\end{figure}

\subsubsection{Gravity Waves in the Solar Tachocline}

Another possible mechanism for rapid transport of information in the convective interior could come in the form of {\revone gravity waves} in the solar tachocline \citep{1992A&A...265..106S} \-- the shear layer between the Sun's radiative interior and convection zone (see Figure~\ref{f12}). This idea was explored theoretically by \citet{2019NatSR...9.2035D} which was motivated by the present work. In that work, the gravity waves considered are ``shallow water'' gravity waves, in the sense that they extend across the entire tachocline and manifest in deformations of the top surface. Since shallow-water models do not have density variation with the thickness of the shallow fluid shell, the shallow-water gravity waves are not pure internal gravity waves. The ``effective gravity'' parameter [$G_{\rm eff}$]; a parameter to characterize the shallow-water model) contains within it the Brunt-V\"ais\"al\"a frequency characteristic of the tachocline arising from its subadiabatic stratification of the entire fluid layer. Therefore, they are ``internal gravity waves'' in that sense. The restoring force is in effect felt at all depths, not just at the surface as in pure ``surface gravity waves''. Such waves, like those short-length waves we may observe from a boat in deep water, do not extend very far below the water surface \-- certainly not to the bottom, as shallow-water gravity waves do. The restoring force for surface gravity waves comes from the deformation of the surface only, represented by ``full'' gravity, not just the ``effective'' gravity.

The propagation speed of gravity waves in the solar tachocline \citep{1998ApJ...498L.169B} is dependent on the thickness of the layer, the sun's gravity at that depth, and departures of the stratification from the adiabatic state, [$\delta$], such that $v_{gw} = 1.23 \times 10^{7}~\delta^{1/2} \mbox{ cm s}^{-1}$. Typical values of $\delta$ of the top of the tachocline ($\approx$0.001) and at the center of the tachocline ($\approx$0.04) would respectively yield travel times from the Equator to mid-latitudes of around four days and four hours, respectively, depending on where the wave is triggered. A sudden cancellation of toroidal fields of opposite signs on either side of the Equator (through magnetic reconnection) could excite such a gravity wave because the top boundary of the tachocline would quickly sink down in response to the loss of magnetic pressure. This depression would propagate toward higher latitudes rapidly perturbing the activity bands at mid and high latitudes. This last example would take the form of a ``buoyancy'' wave on the solar tachocline which could conceivably explain the rapid magnetic flux emergence at mid-latitudes {\em and} high-latitudes in a fraction of a solar rotation.

\subsection{The Next Terminator}
The cyclic and somewhat predictable behavior of the magnetic-activity bands \citep{2017FrASS...4....4M} point to a terminator occurring in late 2019 or early 2020 when the activity bands of Solar Cycle 24 reach the Equator, and eventually terminate. This event should permit the growth of activity in what will become Solar Cycle 25 (see, Figure~\ref{f1}C) and launch the rush to the poles that will eventually seed Solar Cycle 26. We note that the Parker Solar Probe will be ideally suited to observe this event in interplanetary space with {\em STEREO-A} while assets on the Sun-Earth line, like \sdo{}, the {\em Interface Region Imaging Spectrograph} \citep[{\em IRIS};][]{2014SoPh..289.2733D} and the Daniel K. Inouye Solar Telescope \citep[DKIST;][]{2014SPIE.9147E..07E}) will allow us to probe the possible excitation and transport mechanisms at play.

\section{Conclusion}
The observations presented here would appear to support a regime where the magnetic fields present in the Sun's convection zone interact with each other at all times, or could potentially be dynamically important. This result gives strong support to the idea that the manifestation of the Sun's large-scale activity, such as the Solar Cycle, is heavily modulated by the interaction and communication of the magnetic systems present in the convection zone. Furthermore, the potential of a terminator event in the next couple of years should lend itself to extensive observational study by the vast array of assets at our disposal, in addition to the array of cutting-edge numerical models attempting to explain their occurrence and phasing in increasingly realistic environments.

\vspace{2em}
This work is dedicated to the memory of Michael J. Thompson \-- scientist, leader, mentor, colleague and friend. Special thanks to Dipankar Bannerjee, Ed Cliver, Subhamoy Chatterjee, Abhishek Srivastava, Ian Hewins, and many others for providing feedback on the material presented. This material is based upon work supported by the National Center for Atmospheric Research, which is a major facility sponsored by the National Science Foundation under Cooperative Agreement No. 1852977. The compilation of feature databases used was supported by NASA grant NNX08AU30G. We acknowledge support from Indo-US (IUSSTF) Joint Networked R\&D Center IUSSTF-JC-011-2016.

\section*{Disclosure of Potential Conflicts of Interest}
The authors indicate that they have no conflicts of interest.

\bibliographystyle{spr-mp-sola}

\end{document}